\documentclass[12pt,a4paper,final]{iopart}

\usepackage{iopams}  
\usepackage{graphicx}

\usepackage{harvard}
\usepackage{color}
\usepackage{url}
\usepackage{pdflscape}
\usepackage{subcaption}

\newcommand{\csflsq}[2]{\chi_{#1}^{(N)\mathrm{#2}}}
\newcommand{\xtotN}{\mathbf{x}_1,...,\mathbf{x}_{N}}
\newcommand{\xptotN}{\mathbf{x'}_1,...,\mathbf{x'}_{N}}
\newcommand{\psik}[2]{\psi_{#1}^{N\mathrm{#2}}}
\newcommand{\ctorb}[1]{\eta_{#1}}
\newcommand{\asymop}{\mathcal{A}}

\newcommand{\xtotNpmom}{\mathbf{x}_1,...,\mathbf{x}_{N-1}}

\begin{document}

\title[Electron correlations in re-collision]{Electron correlations and pre-collision in the re-collision picture of high harmonic generation}

\author{Zden\v{e}k Ma\v{s}\'{i}n$^{1}$, Alex G. Harvey$^{1}$, Michael Spanner$^{2}$, Serguei Patchkovskii$^{1}$, Misha Ivanov$^{1}$, and Olga Smirnova$^{1}$}
\address{$^{1}$Max-Born-Institut, Max-Born-Str. 2A, 12489 Berlin, Germany}
\address{$^{2}$National Research Council of Canada, Ottawa, Ontario, K1A 0R6, Canada}
\ead{olga.smirnova@mbi-berlin.de}

\begin{abstract}
We discuss the seminal three-step model and the recollision 
picture in the context of high harmonic generation 
in molecules. In particular, we stress the importance
of multi-electron correlation during the first and the third of the three steps of the process: (1) the  strong field ionization and (3) the recombination. We point out how accurate account for multi-electron correlations during the third, recombination, step allows one to gauge
the importance of pre-collision: the term coined by 
J. H. Eberly to describe unusual pathways
during the first, ionization, step.
\end{abstract}

\pacs{42.10}
\vspace{2pc}
\noindent{\it Keywords}: High harmonic generation, R-matrix, photoionization, electron correlation.

\submitto{\JPB}

\section{Introduction}

Highly non-perturbative interaction of atoms (and solids) with intense low-frequency laser fields has developed into a subject of intense theoretical and experimental studies already in 1960s, with  the advent of the laser, see e.g.  \citeasnoun{popruzhenko2014} for the review of
both historical and theoretical aspects.
As both the quantity and the quality of the theoretical, especially analytical, models grew, so did their complexity, leaving the underlying physical picture mired in the dark and theorists (at least some) exasperated. Perhaps the first rays
of light reached into the darkness 
with the papers of 
\citeasnoun{kuchiev1987atomic} and 
\citeasnoun{brunel1987not}, showing the first sprouts
of what would later grow into the three-step model.

Soon after,   \citeasnoun{gallagher1988above} 
and \citeasnoun{corkum1989above} 
announced the two step model to the world, boldly forgoing the complexities of quantum mechanics (together with 
\citeasnoun{brunel1987not}) for clarity and 
simplicity of almost classical description of strong-field ionization. The development of the classical three-step re-collision model of atomic interaction with intense infrared light  by
\citeasnoun{corkum1993plasma} and \citeasnoun{schafer1993above}
was aptly qualified by 
\citeasnoun{becker1995light} as "light at the end of the tunnel". 

The three-step model of (1) tunnel ionization in an intense low-frequency laser 
field followed by (2) laser-driven electron oscillations and (3) its return to the parent ion 
provided the unifying picture of strong-field 
phenomena, such as above threshold ionization, high harmonic generation, and highly efficient multiple ionization.
The picture was simple and physically transparent, giving everyone a chance to think how to control and
use this simple dynamics creatively (see e.g.
\citeasnoun{ivanov1995routes}). 
Yet, it did not stop one of the authors of the present paper from complaining "how could the strong 
field physics degenerate to such a level!", mourning 
the loss of obligatory Green's functions and 
multi-dimensional integrals. The goal of
this paper is to avenge this loss and to restore the complexity of strong-field 
theory to its rightful place.

While there still might be only three steps,  each
should be taken with care.

This paper is organized as follows. In Section~\ref{Sec1} we describe the general mechanism of pre-collision in strong-field ionization and in Section~\ref{Sec2} we quantify its importance for the case of CO$_2$ molecule. In Section~\ref{recombination} we describe the molecular R-matrix method and characterize the effects of correlation in the recombination step of HHG from CO$_2$. Finally, in Section~\ref{SecHHS} we discuss how accurate understanding of the effects of correlation in the recombination step allows us to unravel the complexity of the pre-collision.

\section{Multiple channels and the pre-collision in the re-collision model}\label{Sec1}

The basic picture of strong-field ionization in low-frequency fields as  adiabatic tunnelling of the weakest bound, single active electron has long dominated the three step model. In this perspective, the contribution of individual orbitals to ionization is exponentially dependent on their ionization potentials ($I_{p}$). Thus, it is hardly surprising that tunneling was expected to be dominated by electron ejection from the highest-lying molecular orbital (HOMO), with exponentially smaller contributions from a few lower-lying orbitals, which would give rise to additional but hopefully negligible ionization channels. These hopes were dashed  with high harmonic generation experiments in CO$_2$ \cite{kanai2005,vozzi2005controlling,smirnova2009high}. 

High harmonic generation set-up is particularly sensitive to multi-channel 
dynamics. One can easily cope with changing the ionization 
potential from $I_p$ to $I_p+\Delta I_p$ in calculating the ionization amplitudes. 
However, when all channels converge into
their coherent superposition, as they do in high harmonic generation, the 
situation becomes  a lot more - in fact, interferometrically more - complex. Indeed, not only the amplitudes, but also
the phases of the different ionization-recombination channels matter in this 
coherent superposition, forcing one
to think about such concepts as the phase of strong-field ionization and the coupling between different 
ionization channels during strong-field ionization.

\begin{figure}[!htb]
 \centering
 \includegraphics[width=\textwidth]{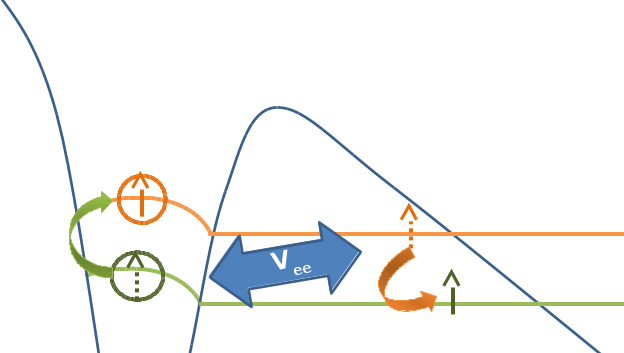}
  \caption{\label{channels} Schematic of the direct and the correlation-induced strong field ionization channels. In direct channels the state of the ion does not change during tunneling while in the correlation-induced channels the state of both the ion and the electron change. The correlation-induced transition can occur both inside and outside the tunnelling barrier.}
\end{figure}

The classical analogue of this coupling has 
been termed pre-collision \cite{pfeiffer2011breakdown,eberly}.
It refers to the 
particular pathway in strong-field ionization  where 
the ejection of one electron is accompanied by the 
excitation of the ion via the interaction of
the departing electron with the core, see 
the sketch in Fig.\ref{channels}. Such ionization pathways are well-known 
in one-photon ionization as shake-up and two-step-one processes. 
In the early days of the 
two-step and three-step models the shake-up has been considered \cite{fittinghoff1992observation}, and rejected
\cite{dietrich1994high}, as a cause of highly 
efficient double ionization in intense low-frequency 
fields. Yet, the shake-up in tunnelling proved resilient, 
re-surging in experiments by 
\citeasnoun{litvinyuk2005shakeup} and \citeasnoun{pfeiffer2011breakdown} and 
in the theoretical analysis (see e.g. \cite{pfeiffer2011breakdown,walters2010,torlina2,pisanty2014momentum,ruberti2018}). Particularly relevant for the present work is the recent theoretical study of~\citeasnoun{ruberti2018} who showed that in aligned CO$_2$ electron correlation modifies the multichannel sub-cycle electron dynamics induced in the parent ion by the strong IR ($800$~nm) ionizing field, interplaying with the laser-driven ionic dipole
interchannel couplings and effectively opening new ionization-recombination channels in HHG. We will show in Section~\ref{SecHHS} that in CO$_2$ the channel couplings are present even in the Mid-IR regime where the coupling due to laser field can be neglected thus exposing the electron correlation built into the system.

Even if the correlation-induced channels are relatively weak compared to
the main ionization channels, which do not
involve ionic excitation by the departing electron, the interferometric nature of the 
HHG signal emphasizes the presence of weak channels,
making them visible in the harmonic signal.

The same sensitivity to multi-electron
dynamics arises in the recombination
step of high harmonic emission. The phase of the 
recombination amplitudes and the coupling between different
recombination channels become very important. Since one-photon recombination is directly linked to one-photon ionization,
and the re-collision in general is directly linked to
electron - ion collision, high harmonic generation brings the full complexity of one-photon ionization and electron - ion collisions \cite{amusia2013atomic,burke2011} squarely 
into the view of strong-field dynamics.

It seems fitting to mark the quarter-century anniversary of the re-collision model by focusing on the multi-electron dynamics of pre-collison and re-collision. 
The three step model of  high harmonic generation not only 
suffers from the complexity associated with
multi-channel dynamics, it also offers routes
to uncovering these dynamics from experiments. 
In particular, interferometric sensitivity of the HHG signal allows one to detect and 
identify the presence of even weak channels in the first, strong-field ionization, step, provided the other two steps are accurately described. 

\section{Pre-collision in strong-field ionization of a CO$_2$ molecule}\label{Sec2}

The Coulomb interaction between the departing electron and the ion left behind can lead to the appearance of the correlation-induced contribution to the ionization amplitude via the process shown in Fig.~\ref{channels}. We will  use the analytical R-matrix method (ARM) \cite{torlina1,torlina2} 
to model this channel and compare the results with ab-initio simulations using the method developed by \citeasnoun{spanner2009}. 

The full details of the ARM approach have been published in~\cite{torlina1,torlina2} and will not be repeated here. The basic principle of the method lies in dividing the space into two parts: the inner region and the outer region. These regions  
are separated by a sphere lying inside the tunnelling barrier. At the matching boundary of the sphere the wavefunction can be well approximated by the wavefunction of the initial neutral state (possibly including its quasi-static polarization by the laser field). Formulating the outer region problem as a boundary-value problem then leads to the expression for the outer region (ionized) part of the wavefunction in terms of an integral over the product of the Green's function (the propagator) for the outer region Schr\"{o}dinger equation with the source term: the neutral state wavefunction projected on the boundary. Resolving the total wavefunction into its channel components, treating the electron correlations to first order and evaluating the integrals using the saddle point method leads to the following expressions for the direct and the correlation-induced ionization amplitudes~\cite{torlina2}:
\numparts
\begin{eqnarray}\label{amps}
a_{nn}^{direct}(\mathbf{p},t_{0}) &=& a_{g}(t_{s})b_{n}(t_{0},t_{s})\exp(-\imath E_{n}t_{0})a_{SAE}(\mathbf{p};I_{p,n},t_{0}),\\
a_{mn}^{corr}(\mathbf{p},t_{0}) &=& c_{m,n}(t_{0},t_{s})b_{n}^{-1}(t_{0},t_{s})a_{nn}^{direct}(\mathbf{p},t_{0}),\label{corramp}
\end{eqnarray}
\endnumparts

where $n$ and $m$ are indices corresponding to the final ionic states which are quasistatic and follow the field adiabatically and $E_{n}$ is the energy of the $n$-th field-free ionic state. 

In these equations, $\mathbf{p}$ is the momentum of the ionized electron and $t_{0}$ is the real part of the complex ionization time $t_{s}$. The functions $a_{g}(t_{s})$ and $b_{n}(t_{0},t_{s})$ represent the Stark shifts of the neutral and the ionic state respectively. We can see that the direct amplitude is proportional to the amplitude $a_{SAE}$ for ionization in the single active electron approximation, which, in turn, depends on the field-free ionization potential $I_{p,n}$.

The correlation-induced amplitude is directly proportional to the direct amplitude. The key quantity determining its magnitude relative to the direct one is the amplitude $c_{m,n}$. In the simplest approximation, when one neglects the Stark shifts in the ion and the difference between the core potentials we get~\cite{torlina2}:
\begin{equation}\label{amp}
c_{mn} = \int_{\tau_{T}}^{0} d\xi \exp(-\Delta I_{p}\xi)\langle m\vert V_{ee}^{n}(\mathbf{r}_{s}(\xi))\vert n\rangle.
\end{equation}
The integral is taken along the line starting from the positive value of the imaginary part $\tau_{T}$ of the complex ionization time down to the real axis (and could be continued further along the 
real time axis to include correlation-driven 
excitation after tunnelling, see \cite{litvinyuk2005shakeup}). 
The trajectory $\mathbf{r}_{s}(\xi)$ of the liberated electron is determined from the saddle-point equations. 
One can see that the exponential factor suppresses 
the contributions of large $\xi$ in favour of those close to the real axis, i.e. close to $t_{0}$ and the exit of the barrier. This is precisely the point illustrated in Fig.\ref{channels}, which 
shows that the correlation-induced transition happens mostly in the vicinity of the tunnel exit. The formula~(\ref{amp}) also shows that the contribution of correlation is not subject to the full exponential penalty associated with the direct channels~\cite{walters2010}.

\begin{figure}[!htb]
 \centering
 \includegraphics[width=\textwidth]{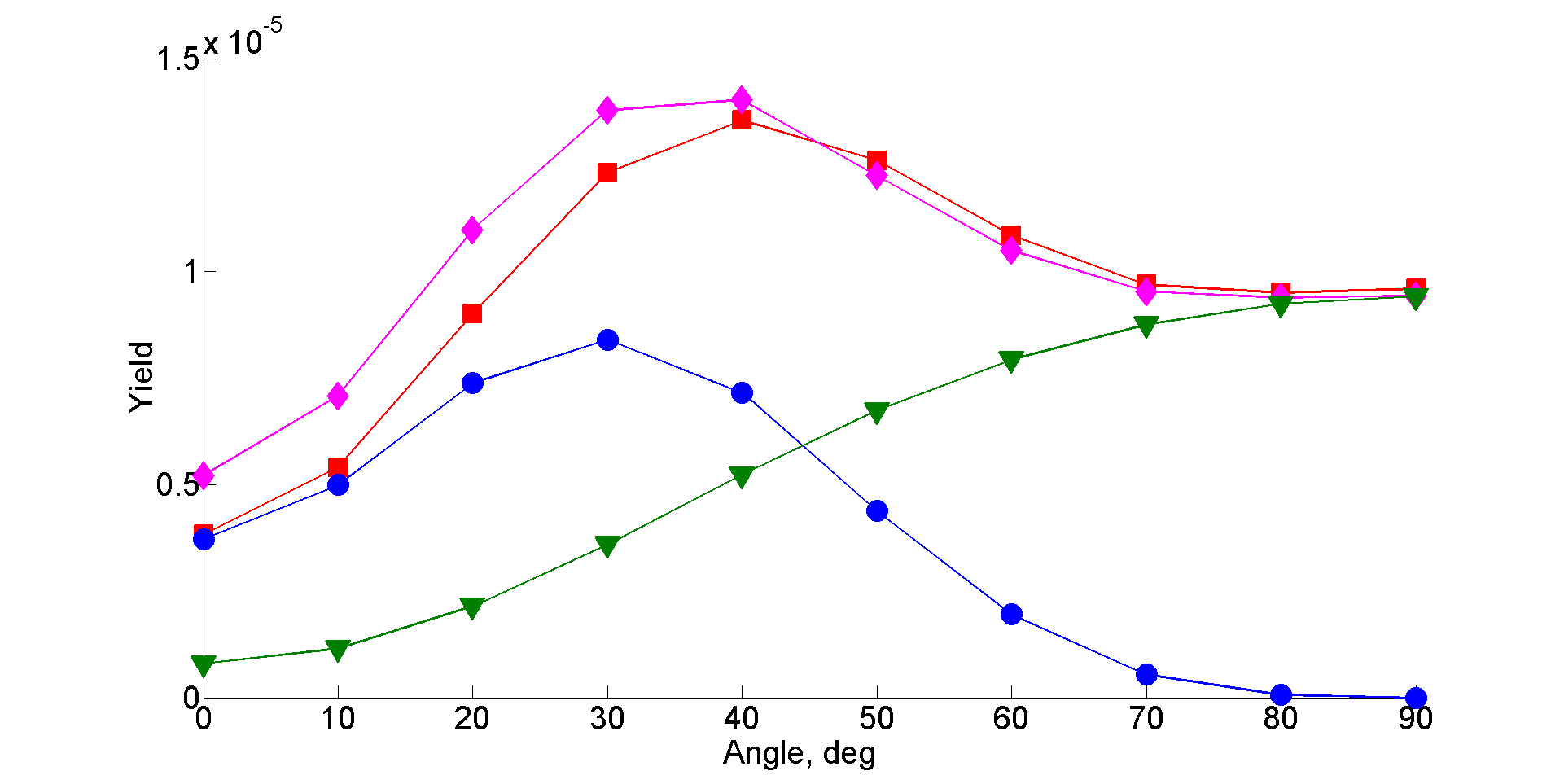}
 \caption{\label{armvsnum} Ionization yields for CO$_2$ molecule obtained for an $800$~nm, $I=1.3\times 10^{14}$ W/cm$^2$ half-cycle pulse, leaving the molecule in the first excited electronic state.  The alignment angle is between the linearly polarized field and the molecular axis.
 Green inverted triangles show results of numerical simulations which do not include the coupling
 between the ionization channels. Red squares show the same 
 calculation which includes the coupling between the 
 channels. The blue circles show the  ARM calculations
 for the correlation-assisted contribution.  Magenta diamonds
 show the ionization yield obtained by adding yields for
 direct and correlation-assisted tunnelling contributions.}
\end{figure}

As an illustration of the potential importance of pre-collision, we first calculated the $c_{mn}$ coefficient for XA correlation channel in CO$_2$ at the peak of the laser field and then  used direct ionization amplitude of the parent ionization channel XX from numerical simulations to obtain the angular resolved yield  $|a_{mn}^{corr}(\mathbf{p},t_{0})|^2$ for the XA channel. The results are shown in Fig.~\ref{armvsnum} (blue circules).
The numerical calculations were performed using the  method of~\citeasnoun{spanner2009} for  the
aligned CO$_2$ molecule, as a function of the
angle between the molecular axis and the polarization of the ionizing field. The ionization yields were obtained for half-cycle of $800$~nm linearly polarized light of intensity $1.3\times 10^{14}$~W/cm$^{2}$. The same numerical calculations can be used to obtain (a) "direct"  ionization yield for channel AA, when the coupling between the channels is turned off (Fig.~\ref{armvsnum} (green triangles)) and (b) "full"  ionization yield for the channel AA, when the coupling between the channels is included (Fig.~\ref{armvsnum} (red squares)).   The results suggest that the correlation-driven channel XA complements the direct channel to provide the "total" ionization yield: the magenta (diamond) curve, which is a sum of the blue (circles) curve and the green (triangle) curve is very close to the total numerical yield. 

In particular, in both cases, the position of the peak in the total yield lies  in the vicinity of $40$~degrees. This
contrasts with the $90$ degrees peak in the single-channel calculation.The angular dependence  of the correlation driven channel XA (blue circles) allows us to associate the shift away from the maximum at $90^{o}$ with the contribution of
other orbitals through the electron-electron correlation.
 The correlation-induced contribution is large and dominates the yield for low alignment angles. The maximum of
the correlation-induced contribution as a function of 
the molecular alignment
angle follows the maximum in the electronic density
of the HOMO (see Fig.~\ref{dysons}), which is where (in the Koopman's picture) the electron is launched from in the correlation-assisted channel.

Recently, the numerical simulations of 
~\citeasnoun{scrinzi2015} have emphasized the role of anti-symmetrization 
(i.e. electron exchange) in 
strong field ionization. In the case
of a CO$_2$ molecule, ionization leaving the molecular
ion in its ground electronic state (i.e.
ionization from HOMO in the Hartree-Fock Koopman's picture) maximizes 
for molecular alignment angles near $45$~degrees. 
This lies in contrast with the geometry of the 
HOMO, where the electronic density maximizes at
about  $30$~degrees with respect to the molecular axis.
This shift is reminiscent of the shift found 
for the first excited A-state of CO$_2^+$. In the 
ARM theory, it is directly linked to 
the contribution of correlation-assisted channels, for which the angular dependence of the ionization yield
is weighted towards the electronic density of the 
initial orbital.

Interestingly, 
an earlier work of~Amusia~\cite{amusia2009,amusia2013} and Flambaum~\cite{flambaum2009,dzuba1982}
analyzed the role of electron exchange comparing
the long-range behavior of the Hartree vs Hartree-Fock (HF) molecular orbitals, i.e.
at the distances relevant for tunnelling.
This work has shown that, at large distances from the core,
electron exchange suppresses the strong exponential decay of the asymptotic part of the inner shell wavefunctions by adding to it a linear combination of the asymptotic parts of the outer shells which do not decay as strongly. This picture could explain the enhancement of the ionization of the inner shell orbitals when compared with the prediction of the Keldysh theory. According to this picture electron correlation beyond the HF approximation contributes in the asymptotic region too but in higher order~\cite{flambaum2009}. This interpretation is consistent with the term ``dynamic exchange" developed by~\citeasnoun{scrinzi2015}: anti-symmetrization of the initial wavefunction 
opens the possibility for electron swapping between
different orbitals in the asymptotic region. Therefore in the formulation of Amusia, Flambaum and Majety the effect of exchange on tunnel ionization can be consistently interpreted as a simple consequence of the Pauli principle: outside of the spatial extent of the inner orbital electrons from the outer orbitals are less constrained to penetrate into this part of space because it is effectively unoccupied by the inner electron.

In the ARM approach the effect of exchange can be analyzed from the starting expression for the ionization amplitude:
 \begin{eqnarray}
 \label{amparm}
 a_{mn}(\mathbf{p},t) &=& 
 \imath \int d\mathbf{k}\int_{-\infty}^{t}dt'
 \langle \mathbf{p}_{t}^{m} m_{t}\vert\mathcal{A} U^{N}(t,t')\mathcal{A}\vert n_{t'}\mathbf{k}_{t'}^{n}\rangle\times
 \nonumber \\
 &\times&\langle n_{t'}\mathbf{k}_{t'}^{n}\vert\mathcal{A}\hat{L}\vert\Psi_{g}^{N}\rangle a_{g}(t')e^{-\imath E_{g}t'},
 \end{eqnarray}
 where $\mathbf{p}$ is the final electron momentum measured at time $t$ and $\vert\Psi_{g}^{N}\rangle$ is the ground state wave-function. The kets $\vert n_{t'}\rangle$ are the quasi-static ionic states and $\vert\mathbf{k}_{t'}^{n}\rangle$ are the 
 corresponding single electron continua 
 defined only in the outer region $r\geq a$, 
 where $r=a$ is radius of the R-matrix sphere which is chosen to lie inside the tunneling barrier~\cite{torlina1}. The operator $U^{N}(t,t')$ is the outer 
 region propagator and $\mathcal{A}$ is the 
 anti-symmetrization operator. We observe that by construction the ARM amplitude has the form of continuum-continuum coupling. Therefore it can be interpreted in terms of elastic ($m=n$) and inelastic ($m\neq n$) electron-ion scattering processes, i.e. as~\textit{pre-collision}. The inelastic collision which results in excitation of the ion left behind contributes to enhancement of ionization of the inner shell orbitals.
 
The exchange interaction enters the ARM amplitude explicitly twice: in the continuum-continuum matrix element on the first line of eq~(\ref{amparm}) and in the matrix element of the Bloch operator on the second line. Exchange in the continuum-continuum matrix element is negligible since the continuum states are defined only in the outer region and therefore overlap very little with the ionic orbitals. This allows us to omit the anti-symmetrization operator in actual calculations~\cite{torlina2} and the pre-collision picture reduces to electron-ion scattering without exchange. For similar reasons the exchange interaction in the Bloch term is negligible too (see Appendix for a more detailed analysis). However, exchange interaction affects the shape of the ground state wavefunction $\vert\Psi_{g}^{N}\rangle$ at $r=a$ for the reasons explained above. In the ARM theory the shape of the ground state wavefunction on the R-matrix sphere is treated as a boundary condition (i.e. as external data) and is not the subject of calculation. Therefore the exchange effects described by Amusia and Flambaum should be included in the Bloch term.
 
We conclude that from the ARM perspective the ``static" electron exchange between the departing electron and those left in the ion is negligible by construction but the theory is still sensitive to the effects described by Flambaum and Amusia through the shape of the ground state wavefunction $\vert\Psi_{g}^{N}\rangle$ on the R-matrix sphere.
 
\begin{figure*}[t!]
    \centering
    \includegraphics[width=\textwidth]{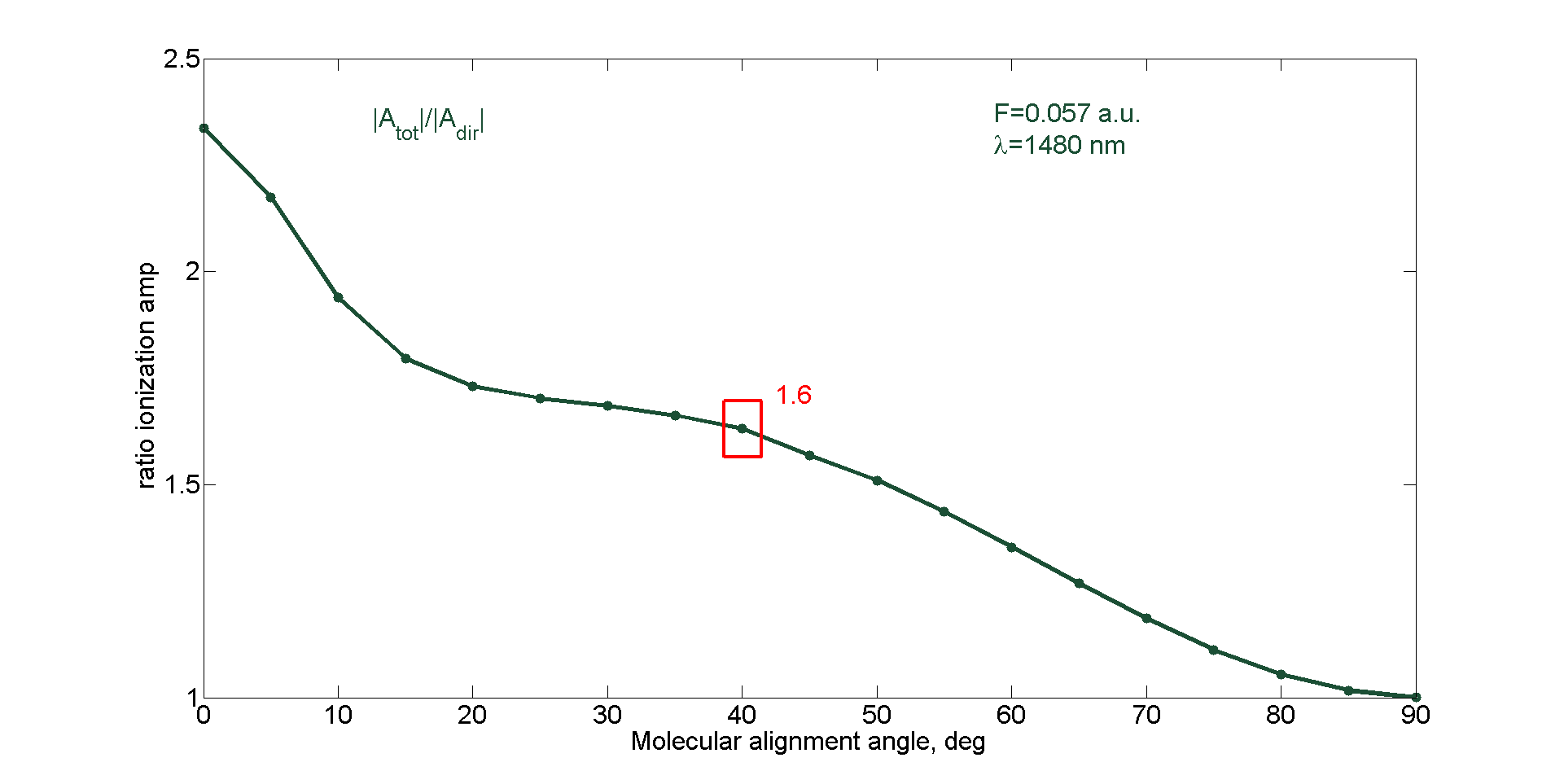}
        \caption{ Ratio of direct vs correlation-induced ionization yields for the A ionic state from the ARM calculations obtained as a function of molecular alignment. The inset shows the value at $40$~degrees.}
        \label{armAstate1480}
\end{figure*}

Recent  experiments~\citeasnoun{bruner2016} have shown that the so-called cross-channels, in which ionization and recombination involve different states of the ion, leave a clear fingerprint in two-color HHG measurements. The correlation-driven transitions during ionization is one of the mechanisms leading to such channels. The other possible mechanism, which is the laser-driven dynamics in the ion between ionization and recombination, is greatly suppressed in the mid-IR fields, which were used in experiments \citeasnoun{bruner2016}.

Our next goal is to analyze the measurements of \citeasnoun{bruner2016} to see if they allows one to identify the contribution of
pre-collision to strong-field ionization into the first excited A-state of CO$_2^+$. The estimates based on the ARM theory show that this contribution should be substantial. 
Figure~\ref{armAstate1480} shows the results of ARM calculations performed at the maximum of the electric field, for $\lambda=1480$ nm and $I\simeq 1.1\times 10^{14}$~W/cm$^{2}$
(as in the experiment by \citeasnoun{bruner2016}).

Importantly, any analysis of the harmonic spectra requires
accurate description of the recombination step. 
In the next section we perform detailed analysis of electron-correlations in recombination  using the ab-initio R-matrix.

\section{Electron correlation in  photo-recombination}\label{recombination}

The effects of correlation in the recombination step are best characterized employing a theoretical approach which allows to control the degree of correlation included in the calculation. For this reason we have used the dipole matrix elements calculated using the UKRmol+ code, a significantly updated version of the UKRmol suite~\cite{ukrmol2012}. The UKRmol and UKRmol+ suites are an implementation of the well-established time-independent molecular R-matrix method with fixed nuclei which has been described in detail before~\cite{tennyson2010,burke2011}. Here we focus only on the essential aspects of the method and its implementation as applied to photoionization (photorecombination). In addition to the standard set of programmes the suite uses the CDENPROP, COMPAK and DIPELM modules~\cite{harvey2014} to calculate the partial wave dipole matrix elements for single photon ionization.

In the previous section we discussed the role of exchange in strong field ionization. The effects of exchange in the recombination step of HHG were studied before~\cite{sukiasyan2010,patchkovskii2006,santra2006} and in the UKRmol suite exchange is included exactly by full antisymmetrization of the multi-electron wavefunction.

\subsection{The molecular R-matrix method}

The R-matrix method divides space into two parts, inner region and outer region, separated by a sphere of radius $r=a$ which must be large enough to fully contain the final (ionic) and the initial (neutral) states of the molecule. The separation of space allows us to solve the Schr\"{o}dinger equation separately in these regions and join them together on the R-matrix sphere using the expression
\begin{equation}\label{rmatdef}
\mathbf{F}(a) = \mathbf{R}_{a}(E)\mathbf{F}^{'}(a),
\end{equation}
where $\mathbf{R}_{a}(E)$ is the R-matrix for total energy $E$ and $\mathbf{F}(a)$, $\mathbf{F}^{'}(a)$ are the one-electron channel reduced radial wavefunctions and their derivatives which describe the motion of the continuum electron in the outer region. Channels are defined as a particular combination of the ionic ($N-1$ electron) wavefunction with the real spherical harmonic from the partial wave expansion of the continuum electron in the outer region. The R-matrix is obtained in the inner region from the spectral decomposition of the Green's function of the Schr\"{o}dinger equation:
\begin{equation}\label{green}
G_{E}(\xtotN ,\xptotN) = \sum_{k}\frac{\psi_{k}^{N}(\xtotN)\psi_{k}^{N}(\xptotN)}{E_{k}-E}, r_{i}\leq a,
\end{equation}
where $\mathbf{x}_{i}$ stands for the space-spin coordinates of $i$-th electron and all space coordinates are restricted to the interior of the sphere. The energy-independent $N$-electron R-matrix basis functions $\psi_{k}^{N}(\xtotN)$ are determined diagonalizing the inner region Hamiltonian matrix:
\begin{equation}\label{innerh}
\langle \psi_{k}^{N}\vert H_{N}+L\vert \psi_{k'}^{N}\rangle = \delta_{k,k'}E_{k}.
\end{equation}
The functions $\psi_{k}^{N}$ are generally non-vanishing on the boundary. The Bloch operator $L$ ensures that (a) $H_{N}+L$ is self-adjoint and (b) that the exact eigenfunctions have the same value of the logarithmic derivative on the boundary. Projecting the Green's function~(\ref{green}) from left and right on a pair of outer region channels and evaluating the resulting expression for $r_{N}\equiv r=a$ gives the elements of the R-matrix:
\begin{eqnarray}\label{rmat}
\mathbf{G}_{E}(r=a,r'=a) = \mathbf{R}_{a}(E) = \frac{1}{2}\mathbf{w}(a)\mathbf{\left[E_{k}-E\right]}^{-1}\mathbf{w}(a)^{T},
\end{eqnarray}
where $\mathbf{w}(a)$ is the matrix of reduced radial amplitudes of the wavefunctions $\psi_{k}^{N}$ in each channel and the matrix $\mathbf{\left[E_{k}-E\right]}^{-1}$ is diagonal. The R-matrix~(\ref{rmat}) determined at $r=a$ can be propagated~\cite{burke2011} to the asymptotic radius $r=c$ where the relationship similar to~(\ref{rmatdef}) together with the known functional form of the asymptotic wavefunctions is used to determine the K-matrix. Transforming the asymptotic standing-wave solution into one satisfying the photoionization or scattering boundary conditions and backpropagating it provides the reduced radial wavefunctions $\mathbf{F}(r)$ and their derivatives on the boundary. The Green's function~(\ref{green}) together with the reduced radial wavefunctions on the boundary are then used to obtain the wavefunction in the inner region:
\begin{equation}
    \Psi_{E}^{N}(\xtotN) = \sum_{k}A_{k}(E)\psi_{k}^{N}(\xtotN), r_{i}\leq a,
\end{equation}
where the coefficients $A_{k}(E)$ depend on the derivatives $\mathbf{F}^{'}(a)$, see~\citeasnoun{harvey2014} for the full equations. This completes the determination of the wavefunction in all space.

The construction of the inner-region eigenfunctions $\psi_{k}^{N}(\xtotN)$ is the most important and the most difficult part of the calculation since their form determines the level of modelling of the electron-electron interaction and the number of ionic channels included. These functions have the form of Close-Coupling expansion:
\begin{eqnarray}\label{eqclosecoup}
\psik{k}{}(\xtotN)&=&\asymop \sum_{i,j} a_{kij}\phi_{i}^{N-1}(\xtotNpmom)\ctorb{ij}(\mathbf{x}_{N})+\nonumber \\
&+&\sum_{p} b_{kp} \csflsq{p}{}(\xtotN).
\end{eqnarray}
The configurations $\csflsq{p}{}$ are built from the molecular orbitals which are fully contained inside the R-matrix sphere and describe correlation/polarization of the molecule in a manner similar to the standard configuration-interaction method from quantum chemistry. To describe the possibility of electron ejection to the outer region the configurations listed on the first line of eq (\ref{eqclosecoup}) are included. These configurations are non-zero on the boundary and are constructed coupling the ionic wavefunctions $\phi_{i}^{N-1}$ to the continuum spin-orbitals $\ctorb{ij}$ and anti-symmetrizing the result. Finally, the coefficients $a_{kij}$ and $b_{kp}$ are determined diagonalizing the inner-region Hamiltonian as required by equation~(\ref{innerh}).

The unique capability of the UKRmol implementation lies in its flexibility in describing electron correlation in the inner region on levels ranging from the simple Hartree-Fock (HF) approach to the Close-Coupling which includes multiple correlated states of the ion in eq (\ref{eqclosecoup}) and a sophisticated modelling of the correlation/polarization effects using complete-active-space configuration-interaction (CAS-CI). For the purpose of connecting the photoionization calculations with modelling of HHG it is also convenient that the physical picture underlying the ab initio R-matrix method closely resembles that of the three-step model of HHG which separates the complicated dynamics in the inner region (ionization and recombination) from the simple intermediate step (propagation in the continuum). For the CO$_2$ photoionization calculations we have used R-matrix sphere with radius of $10$~Bohr.

\subsection{Requirements on the photorecombination dipoles}

In the context of the calculations for CO$_2$ the capability to describe accurately multielectron effects has been shown to be important for accurate description of the partial single-photon ionization cross sections for the ground and the first three excited states of the ion, especially the C state~\cite{harvey2014}. The need for highly accurate dipole matrix elements becomes especially important when we consider the specific requirements on their quality which arise when these are included in the three-step model of HHG.

\begin{enumerate}
    \item The interferometric property of HHG demands a balanced and accurate description of the photorecombination step from the number of different participating ionic channels: inaccuracies in the phases of the dipole matrix elements, their relative strengths and/or angular dependencies as a function of photon energy all combine to produce distortion of the calculated HHG spectrum. If we seek to reconstruct the dynamics between ionization and recombination from the experimental data then this can be significantly affected by the distortion of the calculated HHG spectrum due to these inaccuracies. The redistribution of the returning electron flux into the individual ionic channels is controlled by continuum-continuum coupling, i.e. by polarization and real electronic excitation of the ion by the returning electron. The amplitude for recombination (the dipole matrix element) is further affected by the correlation effects in the final neutral state. However, both of these effects are important only inside the inner region: in the outer region the interaction between the recombining electron and the molecular ion is dominated by the static Coulomb interaction. Higher multipoles of the static electron-ion interaction potential as well as dynamical polarization of the ion are typically negligible in comparison.

    \item In HHG driven by a long wavelength (mid-IR) laser-field the energy of the returning electron spans a wide range, in our case up to $\approx$~$80$~eV. While simple approaches such as the Eikonal-Volkov method~\cite{smirnova2008} can be used successfully to calculate the dipole matrix elements in the high-energy range, they typically become very inaccurate in the low-energy part of the spectrum where multi-electron effects take place and specialized methods exist to provide accurate data in this regime. However, both the high and the low energy part of the spectrum are linked, via the energy-time Fourier integral, to the dynamics taking place in the half cycle of the laser field. Therefore a smooth dependency of the dipole matrix elements on energy is required. Consequently, we require that the same level of theory is used throughout. In the low-energy range accurate modelling of polarization/correlation is necessary and can be achieved using the UKRmol suites. However, up to now these codes have been mostly applied to problems where the unbound electron's energy does not significantly exceed the ionization potential of the target molecule, typically around $10$~eV. In the UKRmol suite the radial parts of the continuum orbitals are represented using nodeless Gaussians making it very hard to represent the highly-oscillating wavefunctions of the energetic electrons. For the present application the requested maximum electron energy is almost an order of magnitude larger than the typical electron energies required, demonstrating the magnitude of the computational challenge. Inaccurate description of the continuum wavefunction causes unphysical oscillations of the dipole matrix elements as a function of energy or indeed a complete breakdown of the results.

\end{enumerate}

In the following we show how the first challenge is met in our calculations using sophisticated modelling of electron correlation and the second challenge using our upgraded codes, UKRmol+, which are capable of representing the continuum for high electron energies, in this case up to $\approx 80$~eV.

\subsection{Photoionization models for CO$_2$}

In the previous R-matrix study on CO$_2$~\cite{harvey2014} several R-matrix scattering models were studied in detail. Here we restrict ourselves to a comparison of two models: the HF model and the highest-level CAS-CI model (Model~$3$ in~\citeasnoun{harvey2014}). The correlated CAS-CI model uses active space of $15$~orbitals and $15$~electrons to describe the ionic states $\phi_{i}^{N-1}$ in eq~(\ref{eqclosecoup}), generating approx. $3600$ configurations for each doublet spatial symmetry. The $L^2$ functions $\csflsq{p}{}$ are generated adding one electron to the same active space resulting in approx. $18600$ configurations per each singlet spatial total symmetry. The calculations were carried out in the $D_{2h}$ point group: the UKRmol suites in common with most quantum chemistry software cannot use the actual non-Abelian $D_{\infty h}$ point group of the molecule. 

In the original calculations of~\citeasnoun{harvey2014} $96$~ionic states were included and the largest photon energy was limited to $50$~eV. The calculations presented here differ from the previous ones by including $300$~ionic states spanning the energy range of approx. $30$~eV and by extending the photon energy from $50$~eV to $90$~eV. Extension of the energy range for the unbound electron required increasing the largest continuum angular momentum from $l=5$ to $l=7$. We are interested in studying the photoionization observables only for the four lowest-lying ionic states. Including in the calculation a larger number of ionic states than strictly needed generally achieves two things: improvement of description of polarization/correlation and mitigation of unphysical pseudoresonances~\cite{tennyson2010}. In the original calculations polarization/correlation was already well converged and therefore in the present calculations the additional states mostly contribute to elimination of the pseudoresonances. However, tripling the number of ionic states included increases significantly the computational demand, especially in the outer region: for high energies the number of open channels in each singlet spatial symmetry reached around $2400$. These demanding calculations have been turned into routine ones following a recent MPI parallelization of the outer region code.

To represent the continuum accurately in the new calculations we used the upgraded UKRmol+ suite which is based on a completely new parallel integral core capable of producing multi-centric molecular integrals over Gaussian orbitals and mixed integrals between atom-centered Gaussians and center-of-mass centered B-spline orbitals in both double and quad precision. In our calculations the continuum functions centered on the center of mass can be either Gaussians, B-spline orbitals or a combination of both. The choice of the type of the continuum functions depends on the size of the target molecule and on the required energy range for the continuum electron. For the present application R-matrix radius of $10$~Bohr was sufficient. This value is small enough to allow the use of Gaussian-only continuum basis to represent the continuum for energies up to approx. $80$~eV. However, this approach works only if the integrals are calculated in quad precision. The use of quad precision is essential since it allows us to eliminate problems with numerical linear dependencies between the diffuse continuum Gaussian orbitals and to retain all available continuum orbitals in the basis. In some applications using Gaussian-only continuum is not possible even if quad precision is used and the B-spline orbitals must be included~\cite{darby2017}.

\begin{figure}[!htb]
 \centering
 \includegraphics[width=\textwidth]{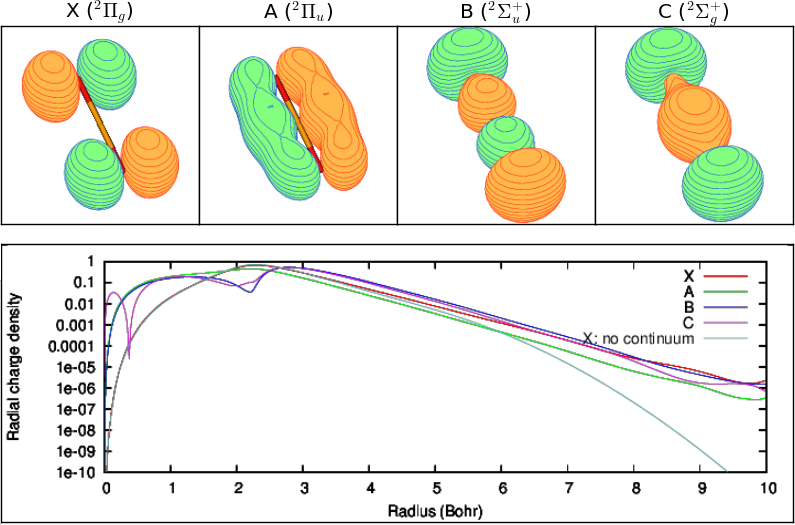} 
 \caption{\label{dysons} Top panel: Dyson orbitals for the four lowest-lying ionic states of CO$_2$ obtained from the CAS-CI R-matrix calculation and displayed for contour value of $0.01$. Bottom panel: radial charge densities of the Dyson orbitals. Dyson orbital for the X state obtained from a calculation not including any continuum orbitals in the neutral state description is also shown. Note the logarithmic scale on the y-axis.}
\end{figure}

In our calculations the lowest-lying R-matrix basis state of the $^{1}A_{g}$ ($^{1}\Sigma_{g}$) symmetry is used as the initial neutral state of the molecule. This state has the form of configuration expansion given by eq (\ref{eqclosecoup}) and therefore includes the contribution of the continuum orbitals. The continuum basis is large and diffuse enough to ensure accurate representation of the monoexponential decay of the neutral bound wavefunction in the asymptotic region close to the R-matrix sphere. Consequently, this property allows us to obtain the correct exponential decay of the Dyson orbitals, see Fig.~\ref{dysons}, which shows the radial charge densities of the Dyson orbitals for the four ionic states as obtained from the R-matrix CAS-CI model.
To demonstrate explicitly the importance of inclusion of the continuum orbitals the figure shows also the radial charge density of the Dyson orbital for the X-state obtained from a calculation where the functions on the first line of eq~(\ref{eqclosecoup}) were removed from the inner region basis for the neutral state. In this case the atom-centered GTO basis stops being able to represent the correct asymptotics already at radius around $6$~Bohr. 

We would like to point out that the Dyson orbitals obtained in this way will necessarily have the property that their logarithmic derivative on the R-matrix sphere converges to zero as the size of the continuum basis approaches completeness. This phenomenon can be seen in the results in Fig.~\ref{dysons}. This is a consequence of two things: (1) the presence of the Bloch operator in the inner region Hamiltonian which ensures that the exact eigenfunctions of the inner region Hamiltonian satisfy the zero logarithmic derivative boundary conditions on the R-matrix sphere and (2) of our choice to use the lowest-lying R-matrix state to approximate the initial neutral state. In practice this issue does not cause problems as long as the R-matrix sphere is chosen large enough to ensure that the condition of zero logarithmic derivative on the boundary lies in the region where the amplitude of the wavefunction is negligible. Nevertheless, it is important to keep this property of the R-matrix theory on mind especially in applications where amplitudes of the Dyson orbitals in the asymptotic region are required. Alternatively, the derivative problem can be avoided completely matching the inner region solutions to proper bound state asymptotics on the R-matrix sphere.

\begin{figure}[!htb]
 \centering
 \includegraphics[width=\textwidth]{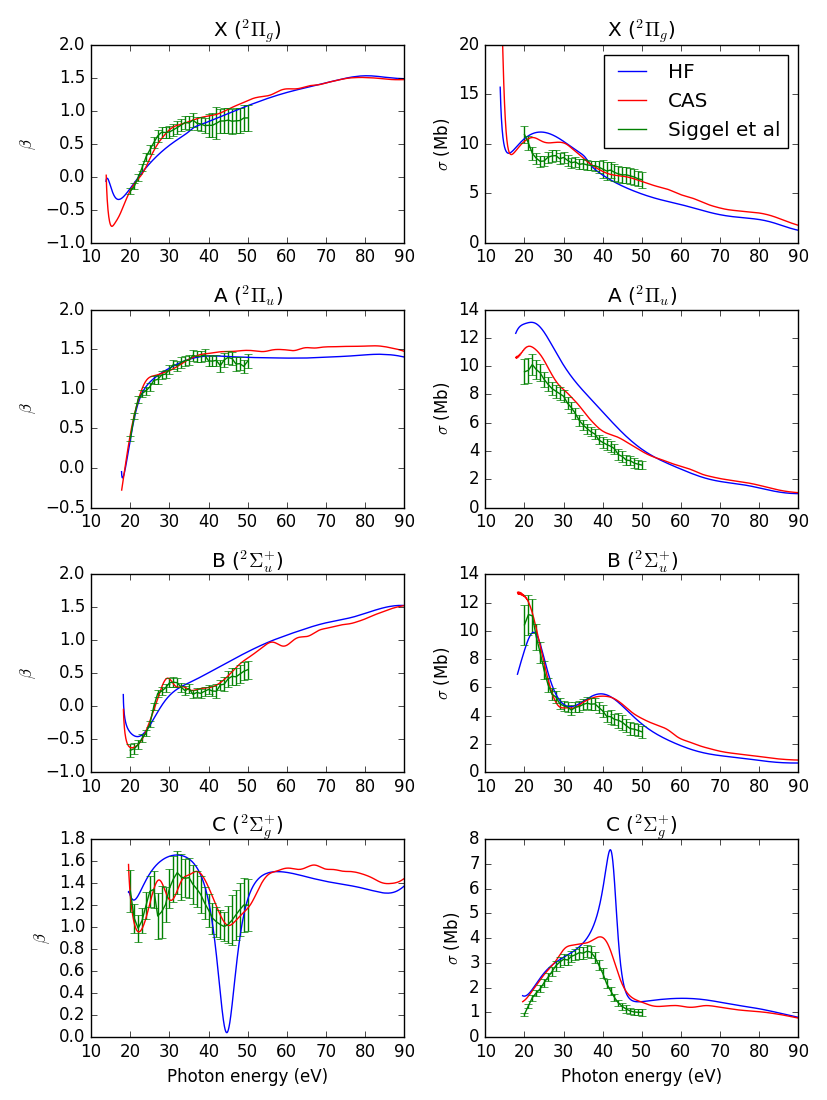}
 \caption{\label{cs_beta} Partial cross sections in Mb (right column) and $\beta$-parameters (left column) for the four ionic states (in rows) as obtained from the HF (blue lines) and CAS-CI (red lines) calculations. The green lines are the experimental values of Siggel et al.}
\end{figure}

\subsection{Results}

In Fig.~\ref{cs_beta} we show the photoionization partial cross sections and $\beta$~parameters for the four lowest-lying ionic states as obtained from the HF and the new CAS-CI model. The CAS-CI dipole matrix elements have been smoothed with a Gaussian (FWHM = 0.05 Hartree) to remove the dense jagged structure associated with the autoionizing resonances. The experimental values of the $\beta$~parameters and the partial cross sections~\cite{siggel} are shown in green. The partial cross sections were obtained normalizing the branching ratios of~\citeasnoun{siggel} to the total cross sections of~\citeasnoun{chan1993}. Clearly, the agreement between the highest-level (CAS-CI) results and the experiment is excellent. Differences between the HF and the CAS-CI results correspond to regions where correlation/polarization plays an important role. As expected, these regions are limited to lower energies below approx. $50$~eV in photon energy. In the higher energy region the HF and the CAS-CI results are in good agreement. In the CAS-CI results additional features appear which are not present in the HF results: the shoulder in the $\beta$~parameter for the X channel around $30$~eV, the rise and fall of the $\beta$~parameter for the B state in the region $25-50$~eV and the large change in both the $\beta$~parameter and the cross section for the C state in the wide energy range between $30-50$~eV. The latter is caused by the well-known wide resonance. This resonance is nominally shape but has a strong inelastic component. Overall we can see that the contribution of correlation/polarization is non-negligible at low photon energies and especially for the C state it plays a major role.

In physical terms the strong inelastic component of the C-channel resonance means that the electron ejected from the $4\sigma_{g}$ orbital temporarily attaches to the resonance which has the property that it strongly enhances transfer of energy, via the Coulomb interaction, between the electron and the ion leading to its excitation/dexcitation and thus diminishing the amplitude for finding the ion in the C-state at the end of the interaction. Apparently, this effect is strongest for photon energies around $40$~eV. A manifestation of this resonance-enhanced inter-channel coupling is probably the small change in the $\beta$~parameter for the B state between $25-50$~eV, see also the discussion in~\citeasnoun{siggel}. The simplest model for the inter-channel coupling is the dipolar interaction between the continua of the two channels. The strength of this interaction is proportional to the transition dipole moment between the corresponding ionic states. The A and B states are both dipole-coupled to the C state but only the B state has a large magnitude of the dipole transition moment with the C state, approx. $1$~a.u. The magnitude of the transition dipole with the A state is only about $0.01$~a.u. Therefore the strong dipolar interaction between the continua of the B and C states can at least partially explain the change of the $\beta$~parameter for the B state when correlation is included. 

Overall, the difference between the HF and the correlated CAS-CI cross sections for the X,A,B states in the vicinity of the resonance are not large enough to explain fully the strong suppression of the resonant peak in the C channel. Convergence of the shape of the resonant peak and the angular distributions is only achieved including a large number of excited ionic states in the Close-Coupling expansion, see eq~(\ref{eqclosecoup}). Our results clearly show that the resonance in the C channel couples predominantly to the highly excited states of the ion including multiple-hole states. As demonstrated earlier~\cite{harvey2014}, our accurate CAS-CI calculations are a major improvement over early few-channel calculations, cf~\citeasnoun{siggel} and references therein. We can see that the ability of the UKRmol suites to include a large number of highly excited states in the photoionization calculation is essential for obtaining accurate observables and dipole matrix elements for CO$_2$.

Note that the broad peak in the cross section for the B state around $40$~eV is not related to the C-channel resonance since it appears already on the HF level where no interchannel-coupling is included. Instead, this peak is a manifestation of non-resonant enhanced ionization into p-wave continuum channels~\cite{siggel}.

\begin{figure}[!htb]
 \centering
 \includegraphics[width=\textwidth]{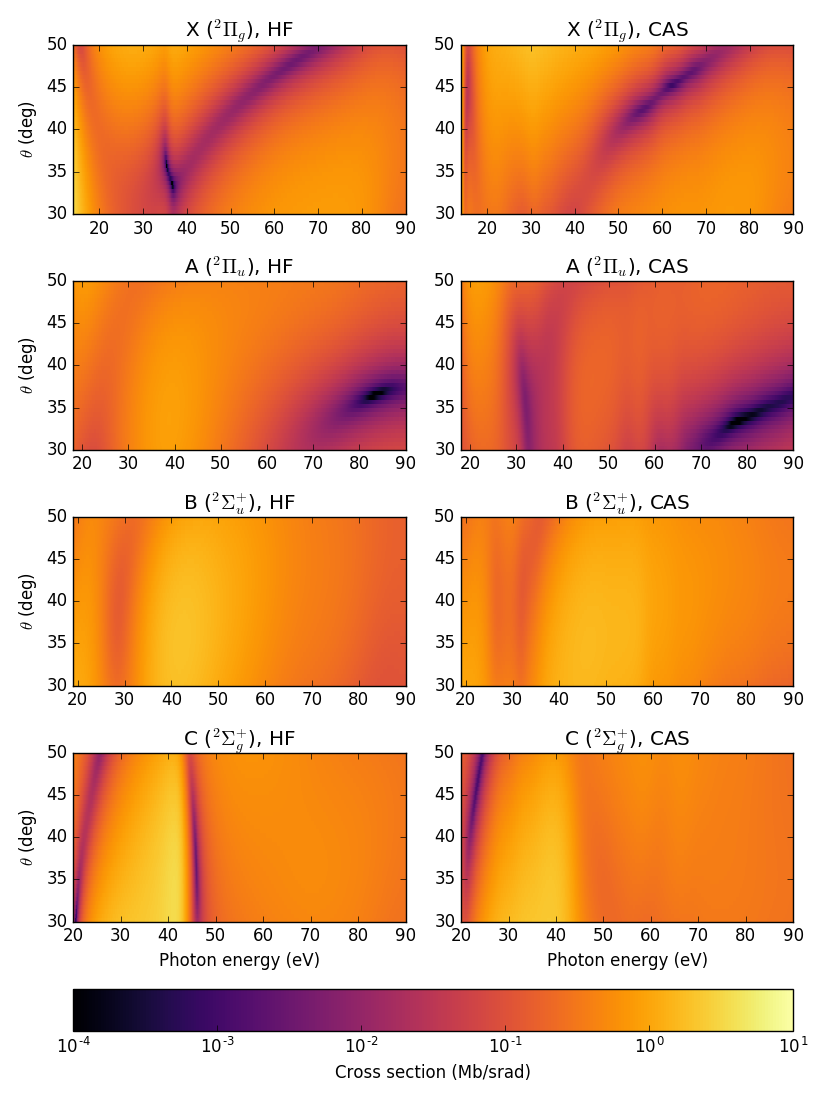}
 \caption{\label{d_dcs}Cross sections for parallel light: HF (left column) vs CAS-CI (right column) results for the four ionic states (in rows). The cross sections are given in Mb/srad (note the logarithmic scale) as a function of the photon energy (in eV) and the alignment angle $\theta$ (in degrees).}
\end{figure}

\begin{figure}[!htb]
 \centering
 \includegraphics[width=\textwidth]{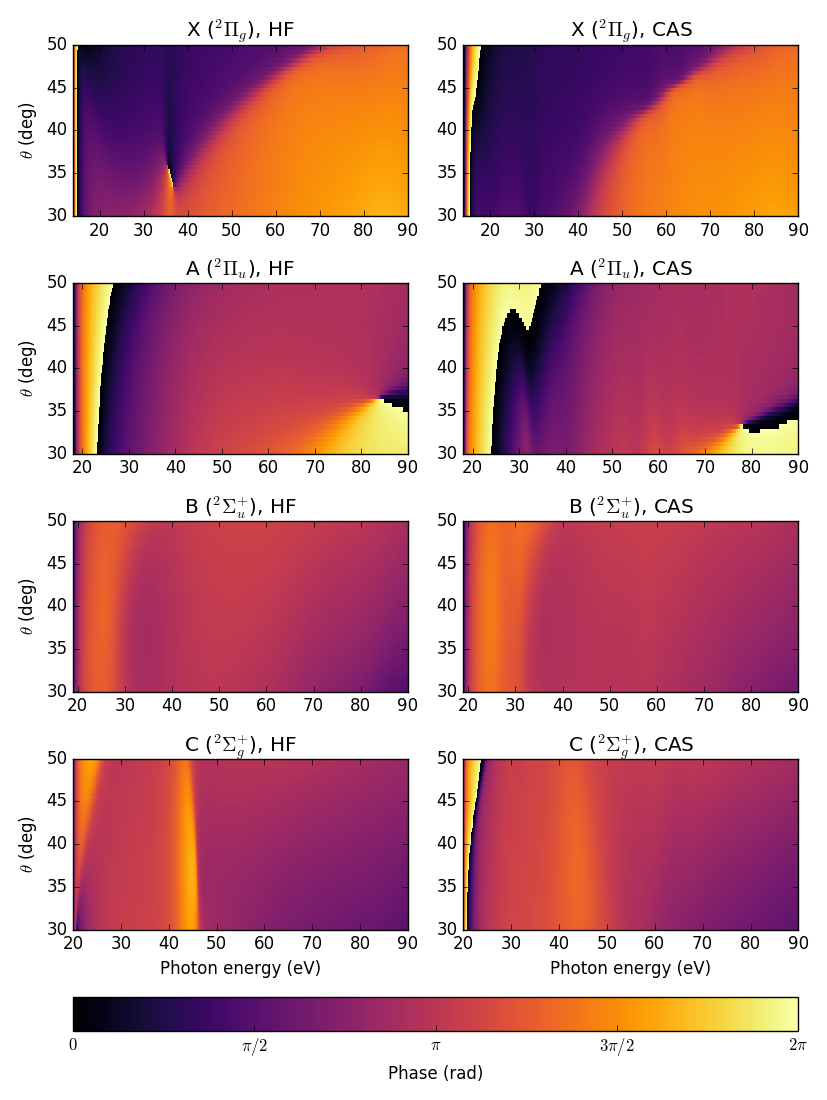}
 \caption{\label{d_phases}Phases of the dipoles for parallel light: HF (left column) vs CAS-CI (right column) results for the four ionic states (in rows). The phases are given in radians as a function of the photon energy (in eV) and the alignment angle $\theta$ (in degrees).}
\end{figure}

Those features in the HHG spectra which do not depend significantly on the parameters of the driving laser field are most-likely associated with structure, i.e. with the recombination step. A very useful diagnostic in this case are the dipole matrix elements and the associated cross sections for parallel light: the emission of a single photon polarized parallel to the direction of the returning electron. The cross sections and phases of the dipoles (modulo $2\pi$) are shown in Figs.~\ref{d_dcs} and~\ref{d_phases} as a function of photon energy and the molecular alignment angle measured wrt direction of photon polarization. We only show the results for alignment angles around $40$~degrees since those are the ones which contribute most strongly to the coherent orientational averaging in HHG: the $\sin(\theta)$ factor in the integral over the alignment distribution suppresses contributions of the smaller alignment angles. We now analyse Figs.~\ref{d_dcs} and~\ref{d_phases} for each of the final states separately. 

For the X state the cross sections in Fig.~\ref{d_dcs} show the well-known structural minimum~\cite{kanai2005} which then appears as a sharp $\pi$-jump in the dipole phase in Fig.~\ref{d_phases}. The structural minimum is caused by interference of amplitudes for recombination into the two components of the Dyson orbital: each component is localized on one oxygen atom, see Fig.~\ref{dysons}. However, thanks to dynamical polarization, the electron incoming under a certain angle wrt molecular axis causes different distortions of the two components of the hole in the ion thus breaking the symmetry between the two recombination amplitudes. The dynamical polarization is important for low energies of the ejected electron and is included in the CAS-CI calculation. It can explain the slight suppression of the structural minimum in the CAS-CI results in comparison with the HF results at energies below $\approx 50$~eV. The sharp feature visible in the HF results around $35$~eV survives even in the CAS-CI calculations but is suppressed and moves down in energy to approx. $30$~eV.

The cross sections and dipole phases for the state A show structural differences below approx. $40$~eV and then again at high energies above approx. $60$~eV. Interestingly, the structure at higher energies and alignment angles below $40$~degrees resembles the structural minimum in the X state. However, in contrast to the X state the structural minimum here is constrained to higher electron energies (reflecting the smaller distance between the two components of the Dyson orbital for the state A) and alignment angles below $40$~degrees. Fine details of this structure depend on the model: the minimum moves to lower alignment angles in the correlated CAS-CI results.

The results for the B state obtained from the two calculations show only quantitative differences while their structure remains very similar. For the C state, in the vicinity of the resonance between $40-50$~eV, both the cross section and the magnitude of the jump in the dipole phase are significantly suppressed in the CAS-CI results. This effect is explained as above using the redistribution of the outgoing electron flux from the C-channel to a large number of highly excited ionic channels.

\begin{figure}[!htb]
 \centering
 \includegraphics[width=\textwidth]{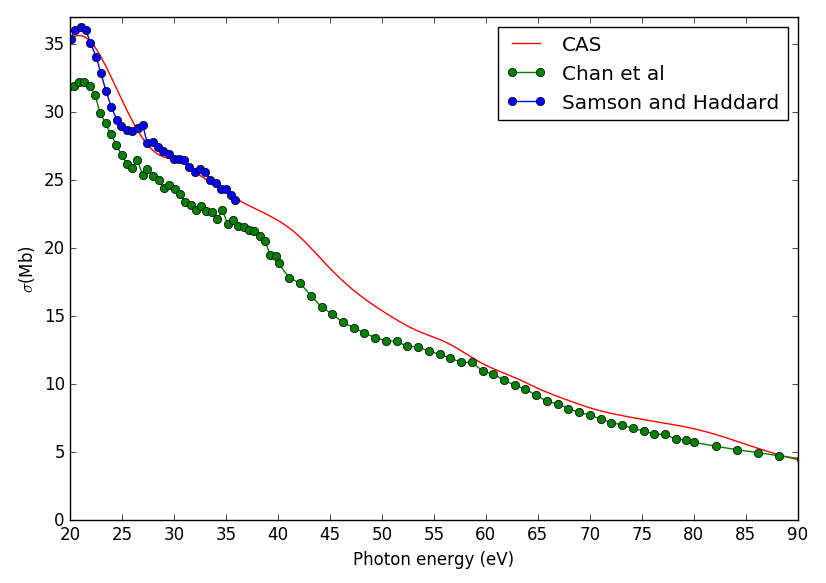}
 \caption{Total cross section for CO$_2$ calculated in the present work on the CAS level and the experimental total cross section of~\protect\citeasnoun{chan1993} and Samson and Haddard~\protect\cite{gallagher}. }\label{total_cs}
\end{figure}

Finally, in Fig.\ref{total_cs} we compare the total cross section calculated on the CAS level with the experimental data of~\citeasnoun{chan1993} and Samson and Haddard as reported in~\cite{gallagher}. The CAS cross section was obtained summing the contributions of the four dominant partial cross sections for the X,A,B,C states. The agreement with the low-energy data of Samson and Haddard is excellent. The data of~\citeasnoun{chan1993} are offset to slightly smaller values but the shape of the cross section remains in an excellent agreement with our theory in the whole energy range.

\begin{figure*}[!htb]
    \centering
    \begin{subfigure}[t]{0.5\textwidth}
        \centering
        \includegraphics[width=\textwidth]{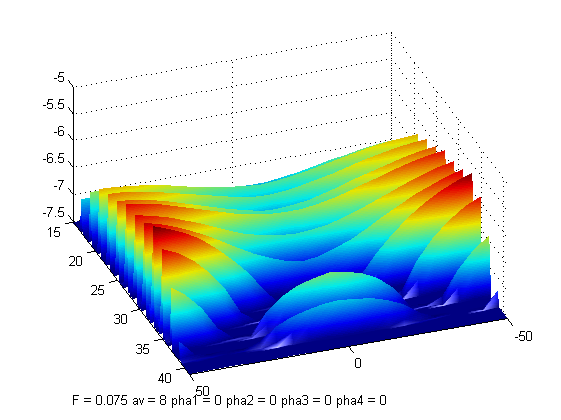}
        \caption{\label{HHGCAS} Correlated CAS-CI photorecombination dipoles.}
    \end{subfigure}%
~
    \begin{subfigure}[t]{0.5\textwidth}
        \centering
        \includegraphics[width=\textwidth]{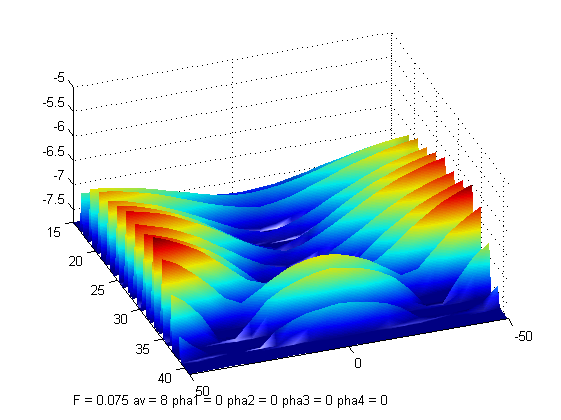}
        \caption{\label{HHGHF} HF photorecombination dipoles.}
    \end{subfigure}
    \caption{\label{HHGcomp} HHG intensity computed using (a) CAS-CI and (b) HF photorecombination dipoles for linearly polarized $800$~nm light with field strength $0.075$~a.u. as a function of the alignment angle. All other parameters were the same in both calculations.}
\end{figure*}

\subsection{HF and CAS-CI dipoles in models of HHG}

To illustrate the differences in the HHG spectra caused by the use of HF and CAS-CI recombination dipoles we show in Fig.~\ref{HHGcomp} the HHG intensity obtained from the full three-step model of CO$_2$ for $800$~nm linearly polarized light of field strength $0.075$~a.u. In both cases all quantities used in the AZUR calculation (see below) were kept the same with exception of the photorecombination dipoles which were taken from the CAS-CI and the HF calculations respectively. These calculations used a simplified model of HHG which included the ionization-recombination diagonal channels XX, AA, BB, CC and the cross-channels XB and XC. The purpose of this calculation was to investigate the effect of the recombination dipoles while still obtaining realistic HHG spectra. The spectrum is shown in the vicinity of the dynamical intereference minimum~\cite{smirnova2009high} as a function of the alignment angle and photon energy. We observe that the HF dipoles predict a much deeper minimum and a slightly different angular dependence of the emission. This is consistent with the analysis of the dipole matrix elements which show more differences at low photon energies, where correlation/polarization is important.

\section{High harmonic spectroscopy of pre-collision.}\label{SecHHS}

Detection of correlation-assisted channels in optical tunnelling is extremely challenging. If photo-electron 
spectroscopy is used, then identification of multiple ionization channels requires one to perform coincidence measurements~\cite{boguslavskiy2012}. 
Strong field ionization into 
different states of the ion  would lead to different 
series of peaks in the photo-electron spectrum 
determined by the energy conservation law, $E_{N,n}=N\hbar \omega -I_{p,n}$, where  $N$ is the number of absorbed photons
and $I_{p,n}$ is the ionization potential associated with
$n$-th ionization channel.
However, this spectrum does not yet distinguish direct tunnelling into excited ionic states from correlation assisted tunnelling into the same states, which happens during the ionization step.

One way to detect pre-collision is to resolve times of
ionization and recombination using two-dimensional High Harmonic Generation 
Spectroscopy~\cite{serbinenko2013,bruner2015,bruner2016}. 
In this setup, the strong linearly polarized driving field 
is complemented with a weak second harmonic,  
orthogonally polarized and phase-delayed with respect to the driving field. The second field is sufficiently weak to minimize its impact on the ionization dynamics.
However, it affects the recollision: its action on the liberated electron depends on the two-colour delay and determines whether it hits or misses the molecule on return. As a result, 
the effect of the second field is also sensitive to the initial conditions 
with which the electron enters the continuum~\cite{shafir2012}. These conditions depend, in turn, on the ionization dynamics. 

As long as the HHG dynamics proceeds through a single channel, 
the response to the second field is universal ~\cite{shafir2012}. However, if electron-electron correlation is present at the ionization step, the universality is lost. The correlation acts on both the ion and the liberated electron, altering the timing and 
the initial momentum distribution with which it appears in the continuum \cite{pisanty2014momentum} and generating cross-channels
in high harmonic emission: the system starts in one channel
and ends in the other, altering timing of the radiative
recombination event for a given harmonic. 

Cross-channels in HHG
can also arise due to transitions induced by the laser field, but in mid-IR fields such transitions are suppressed. The 
field-induced virtual excitation amplitudes, describing quasi-static polarization, will 
evolve with the energy of the original, polarized state rather than the energy of the excited state and can easily be  accounted 
for in the simulations. 
Thus, the presence of pre-collision can lead to real populations
of the excited states and change the response to the 
probe field over a wide range of harmonics. This change  is distinct from 
other effects such as channel interference, which occur in localized parts of the spectrum. Indeed, large deviation of the maximizing two-colour delay from the universal single-channel response was observed in~\cite{bruner2016} over a wide energy range of harmonics between approx.~$50-80$~eV.

In its multi-channel setting, the three step model 
of high harmonic generation turns the 
simple product of the three amplitudes associated
with each of the three steps into a matrix multiplication.
Keeping the convenient  partitioning into the ionization, propagation and recombination steps, in the multi-channel version
the time-dependent dipole has the form ~\cite{smirnova2013,bruner2016}:  
\begin{equation}
\mathbf{D}^{j}(t) = \mathbf{R}^{j}\mathbf{U}^{j}\mathbf{Z}^{j}\mathbf{A}^{j},
\end{equation}
where the index $j$ labels the half-cycles of the driving laser field. 

\begin{figure}[!htb]
 \centering
 \includegraphics[width=0.7\textwidth]{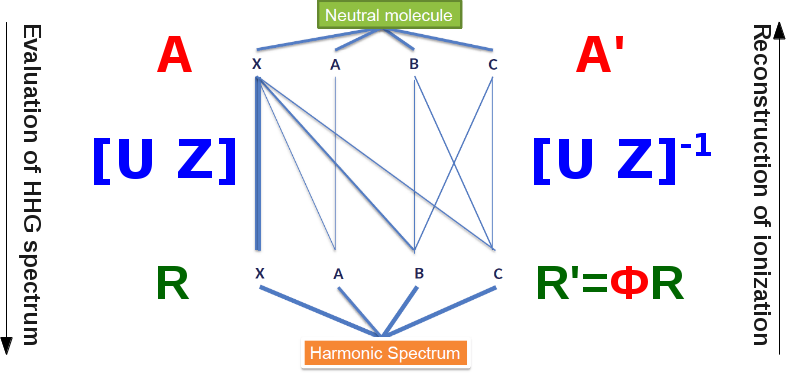}
 \caption{\label{azur} The three-step model in the multi-channel setting.  A is the vector of ionization amplitudes for each of the ionic states involved
 (here four states are sketched). 
 U and Z are the 
 propagator matrices for the ionic states (U) and the continuum electron (Z) when the electron is
 far from the core. R is the vector of dipole recombination matrix elements, which includes electron-electron
 correlation and channel coupling during the recombination step. Some cross-channels are marked by diagonal lines, 
 while the direct channels are marked by the vertical lines. The right hand side illustrates the process of reconstruction of the ionization amplitudes A' starting from the measured two-colour HHG spectrum: the populations in the recombination channels are adjusted by the rotation matrix $\mathbf{\Phi}$ to match the spectrum and back-propagated.}
\end{figure}

Reading from right to left, this AZUR scheme, see Fig.~\ref{azur}, is built from the ionization amplitudes~$\mathbf{A}$, the diagonal matrix~$\mathbf{Z}$ describing propagation of the continuum electron
away from the core, the matrix~$\mathbf{U}$ describing the dynamics in the ion, and the vector of the recombination matrix elements~$\mathbf{R}$. Multiple channels are
coupled only by the laser field during the second step, thanks
to the large distance between the electron and the parent 
ion. However, electron-electron interaction plays important role during the first and the third steps. 
Since the second step is easily modelled,  knowledge of the recombination step allows one to directly connect the
unknown ionization vector $A$ to the observed harmonic
signal, opening the possibility for its reconstruction.

For reconstruction, the initial set of ionization amplitudes 
is taken from single channel theory and propagated using the $[\mathbf{U}.\mathbf{Z}]$ matrices to obtain amplitudes in the recombination channels. These amplitudes are then contracted with the recombination dipole matrix elements obtained from the R-matrix calculations, yielding the induced dipole.  
The resulting induced dipole was Fourier transformed, obtaining the HHG spectrum and the calculation was repeated for a range of two-colour delays determining the 2D HHG spectrum.

\begin{figure*}[t!]
        \centering
        \includegraphics[width=\textwidth]{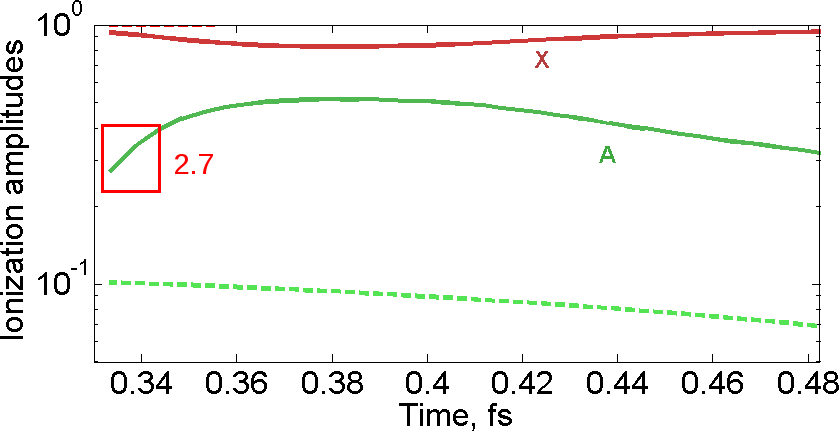}
        \caption{ Reconstructed ionization yields for the X and A states from experimental data. Dashed lines: single channel ionization theory. Solid lines: reconstructed ionization amplitudes. The inset shows the value corresponding to the real part of the ionization time from the three step model.}
        \label{armrec}
\end{figure*}

Since both the modeling of the laser driven dynamics in the ion
and the recombination matrix elements 
are accurate, and since the laser-induced population transfer is 
negligible due to the long wavelength (1480 nm) of the 
driving field, the only source of discrepancy between 
theory and experiment comes from inaccuracy of the ionization vector~$\mathbf{A}$. Furthermore, the propagator matrix $[\mathbf{U}.\mathbf{Z}]$ is unitary and therefore does not change the magnitudes of the populations in the recombination channels. The vector~$\mathbf{A}$ can then be reconstructed from the experimental spectrum adjusting the populations in the recombination channels using a rotation matrix $\mathbf{\Phi}$ so as to obtain agreement with the experimental 2D spectra, see Fig.~\ref{azur}. The resulting  population vectors are then back-propagated using the matrix $[\mathbf{U}.\mathbf{Z}]^{-1}$ obtaining the reconstructed ionization vector~$\mathbf{A}$. For channel A, this procedure yields the results shown in  Fig.~\ref{armrec}. As discussed  in \citeasnoun{bruner2016}, the channel A strongly contributes only to the highest harmonic orders, which correspond to the earliest times in Fig.~\ref{armrec}. Therefore, only the earliest-time results can be reconstructed reliably in this particular case. The ARM estimate of the relative pre-collision amplitude at the peak of the laser field ~$1.6$ is lower but compatible with the reconstructed values $\sim 2.7$ for the short trajectories responsible for the harmonic signal.

\section{Conclusion}

We have studied the role of correlation in the three-step model of high harmonic generation, using the example of the CO$_2$ 
molecule. We combined analytical and numerical approaches and used the experimental results of \citeasnoun{bruner2016} on 2D High Harmonic Spectroscopy in mid-IR ($1480$~nm) fields. 
At these wavelengths, the contribution of inter-channel coupling due to photon absorption alone is suppressed. If present, the inter-channel transitions must be assisted by electron correlation, and can be significant even in the mid-IR regime. 
We presented evidence supporting the presence of 
pre-collision  in strong-field ionization of  molecules
in mid-IR fields.

Our work shows that electron correlation is an important process in strong-field driven processes and can be expected to appear in a large number of molecules. Therefore electron correlation is not limited to the well-known case of  single-photon ionization or recombination, but can also be important during strong-field ionization step.

O.S. acknowledges support from the DFG QUTIF programme SM 292/5 and  the MEDEA project, which has received funding from the European Union’s Horizon 2020 research and innovation programme under the Marie Skłodowska-Curie grant agreement No 641789, MI acknowledges financial support from DFG
and EPSRC/DSTL MURI grant EP/N018680/1.

\section{Appendix}
 Let us analyze the role of 
 anti-symmetrization of the ionic and the continuum states within the ARM theory. 
  , i.e. to the "static" electron exchange between the departing electron and the ion left behind. 
 Consider the starting expression for the ionization amplitude:
 \begin{eqnarray}
 \label{ampfull}
 a_{mn}(\mathbf{p},t) &=& 
 \imath \int d\mathbf{k}\int_{-\infty}^{t}dt'
 \langle \mathbf{p}_{t}^{m} m_{t}\vert\mathcal{A} U^{N}(t,t')\mathcal{A}\vert n_{t'}\mathbf{k}_{t'}^{n}\rangle\times
 \nonumber \\
 &\times&\langle n_{t'}\mathbf{k}_{t'}^{n}\vert\mathcal{A}\hat{L}\vert\Psi_{g}^{N}\rangle a_{g}(t')e^{-\imath E_{g}t'},
 \end{eqnarray}
 where $\mathbf{p}$ is the final electron momentum measured at time $t$ and $\vert\Psi_{g}^{N}\rangle$ is the ground state wave-function. 
 The kets $\vert n_{t'}\rangle$ are the quasi-static ionic states and $\vert\mathbf{k}_{t'}^{n}\rangle$ are the 
 corresponding single electron continua 
 defined only in the outer region $r\geq a$, 
 where $a$ is the radius of the R-matrix sphere. 
 The operator $U^{N}(t,t')$ is the outer 
 region propagator and $\mathcal{A}$ is the 
 anti-symmetrization operator. 

 The matrix element in the first line has the form of a continuum-continuum transition amplitude, where the exchange between the continuum electron and the ionic states is negligible since these states are spatially well separated. However, the matrix element in the second line corresponding to the Bloch operator $L$ is also anti-symmetrized. Let's analyze it in detail. The Bloch operator is defined as follows:
 \begin{eqnarray}
 \hat{L} = -\sum_{j=1}^{N}\hat{\Delta}_{j}(a)\hat{B}_{j} = -\sum_{j=1}^{N}\delta(r_{j}-a)\left(\frac{d}{dr_{j}}-\frac{b_{0}-1}{r_{j}}\right),
 \end{eqnarray}
 where $a$ is the R-matrix radius, $r_{j}$ is the radial coordinate of the $j$-th electron and $b_{0}$ is an arbitrary real constant. 

 Applying the anti-symmetrization in the second matrix element from equation~(\ref{ampfull}) results in an expression for this element, which is a linear combination of the following terms:
 \begin{equation}
   \fl \sum_{j=1}^{N}\langle n_{t}(1,\dots,i-1,i+1,\dots,N) \mathbf{k}_{t}(i)\vert \hat{\Delta}_{j}(a)\hat{B}_{j}\vert\Psi_{g}^{N}(1,\dots,i,\dots,j,\dots,N)\rangle,
\end{equation}
 where the brackets now include indices of the electrons and $i$ is a free index running between $1$ and $N$. For the contribution $i=j$ we first carry out the integration over all electrons except the electron $i$ and obtain:
 \begin{eqnarray}
 \sqrt{N}\langle \mathbf{k}_{t}(i)\vert \hat{\Delta}_{i}(a)\hat{B}_{i}\vert n_{t}^{D}(i)\rangle,
 \end{eqnarray}
 where $\vert n_{t}^{D}\rangle$ is the Dyson orbital for the 
 quasi-static ionic state $\vert n_{t}\rangle$. Therefore this is a direct-type contribution where the $i$-th electron leaving the molecule is projected on the continuum state on the R-matrix sphere. 

 However, the contribution of the $j\neq i$ terms is different: 
\begin{eqnarray}
\langle n_{t}(1,\dots,j,\dots,N) \mathbf{k}_{t}(i)\vert \hat{\Delta}_{j}(a)\hat{B}_{j}\vert\Psi_{g}^{N}(1,\dots,i,\dots,j,\dots,N)\rangle, 
\end{eqnarray}
 and the operator $\hat{\Delta}_{j}(a)\hat{B}_{j}$ now takes the $j$-th bound electron and projects it on the $j$-th electron of the ionic state on the R-matrix sphere while the $i$-th continuum electron is projected on the $i$-th bound electron. Therefore, this is the exchange-type contribution. As long as the R-matrix radius is chosen to lie in the asymptotic region the projection on the $j$-th electron is proportional to the product of two exponentially decaying functions. As a result, for sufficiently large radius $a$, 
 the exchange terms become exponentially smaller in comparison with the direct terms and can be neglected.


\section*{References}
\bibliographystyle{jphysicsB}
\bibliography{bibliography}

\end{document}